\documentclass{aa}

\usepackage{graphicx}
\usepackage{txfonts}
\begin{document}

   \title{A spectroscopic follow-up for Gaia19bld}

   \author{E. Bachelet\inst{1}\fnmsep\thanks{Corresponding author: etibachelet@gmail.com}
            \and
            %Warsaw Gang, order not yet final, to be decided
            P. Zieli{\'n}ski\inst{2}
            \and
            M. Gromadzki\inst{2}
            \and
            I. Gezer\inst{2}
            \and
            K. Rybicki\inst{2}
            \and
            K. Kruszy{\'n}ska\inst{2}
            \and
            N. Ihanec\inst{2}
            \and
            {\L}. Wyrzykowski\inst{2}
            \and
            R. A. Street\inst{1}
            \and
            Y. Tsapras\inst{3}
            \and
            M. Hundertmark\inst{3}
            \and
            A. Cassan\inst{4}
            \and 
            D. Harbeck\inst{1}
            \and
            M. Rabus\inst{1,5}
%       should we add students(A.Gurgul, O.Zi{\'o}lkowska, other who observed it?)
        %    \and
         %   friends\inst{6}
          }

   \institute{Las Cumbres Observatory, 6740 Cortona Drive, Suite 102,93117 Goleta, CA, USA
          %\email{}   
         \and
         Astronomical Observatory, University of Warsaw, 
         Al. Ujazdowskie 4, 00-478 Warszawa, Poland
         %\email{pzielinski@astrouw.edu.pl}
         \and
         Zentrum f{\"u}r Astronomie der Universit{\"a}t Heidelberg, Astronomisches Rechen-Institut, M{\"o}nchhofstr. 12-14, D-69120 Heidelberg, Germany
         \and
         Institut d'Astrophysique de Paris, Sorbonne Universit\'e, CNRS, UMR 7095, 98 bis bd Arago, F-75014 Paris, France 
        \and
             Department of Physics, University of California, Santa Barbara, CA 93106-9530, USA\\
         %\and
         %    Friends Home\\
         %    \email{}
             }

   \date{}
% \abstract{}{}{}{}{} 
% 5 {} token are mandatory
 
  \abstract
   {Due to their scarcity, microlensing events in the Galactic disk are of great interest and high-cadence photometric observations, supplemented by spectroscopic follow-up, are necessary for constraining the physical parameters of the lensing system. In particular, a precise estimate of the source characteristics is required to accurately measure the lens distance and mass.}
   {We conducted a spectroscopic follow-up of microlensing event Gaia19bld to derive the properties of the microlensing source and, ultimately, to estimate the mass and distance of the lens.}
   {We obtained low- and high-resolution spectroscopy from multiple sites around the world during the course of the event. The spectral lines and template matching analysis has led to two independent, consistent characterizations of the source.}
   {We found that the source is a red giant located at $\sim 8.5 $ kpc from the Earth. Combining our results with the photometric analysis has led to a lens mass of $M_l\sim1.1~M_\odot$ at a distance of $D_l \sim 5.5 $ kpc. We did not find any significant blend light in the spectra (with an upper detection limit of $V\le17$ mag), which is in agreement with photometric observations. Therefore, we cannot exclude the possibility that the lens is a main-sequence star. Indeed, we predict in this scenario a lens brightness of $V\sim20$ mag, a value that would make it much fainter than the detection limit.}
  % conclusions heading (optional), leave it empty if necessary % PZ: maybe we should emphasize here shortly the necessity of spectroscopic follow-up of microlensing events in order to determine the parameters and distance to the source and, together with photometric data, lens.
   {}


   \maketitle
%
%________________________________________________________________
%
%-------------------------------------------------------------------
%\onecolumn

\section{Introduction}
The gravitational microlensing method is a powerful way to explore the population of faint objects in the Milky Way. In particular, it has been used over the last 15 years to discover cold exoplanets towards the Galactic Bulge (89 to date, according to the NASA Exoplanet Archive \footnote{\url{https://nexsci.caltech.edu/}}). 
One of the challenges of this method is to obtain an accurate measurement of the lens mass, $M_l$, and distance, $D_l$ \citep{Tsapras2018}. Nowadays, about 50 \% of the published lens systems masses are dependent on galactic models of the Milky Way \citep{Penny2016}. To estimate the mass of the lens, it is necessary to obtain constraints on at least two mass/distance relations. The first relation can be derived by measuring finite source effects in the lightcurve \citep{Witt1994}, parameterized as $\rho$, ultimately leading to the measurement of the size of the angular Einstein ring radius, $\theta_E$ \citep{Yoo2004}. Another route to measuring $\theta_E$ is long-baseline interferometry \citep{Cassan2016}, which up to now, has been achieved for two microlensing events: Kojima-1 \citep{Dong2019} and Gaia19bld (Cassan et al.2021, C21 thereafter). We can also estimate $\theta_E$ by measuring the movement of the light centroid during the course of the microlensing event, referred to as astrometric microlensing \citep{Dominik2000}. This phenomenon should be systematically achievable for events observed by the Gaia space mission \citep{Gaia2016} with $G\le16$ mag and a lens mass $M_l\ge10 M_\odot$ \citep{Lu2016, Rybicki2018}. The measurement of the microlensing parallax, $\pi_E$, provides a second mass mass/distance relation when the source distance $D_s$ is known, since $\pi_{rel} = \pi_E\theta_E$ and $\pi_{rel} = 1/D_l-1/D_s$ \citep{Smith2003}.This can be measured via the "annual parallax" \citep{Gould2004} if the Einstein ring crossing time, $t_E$, is significantly longer than the Earth orbital period or via using the "space parallax," with joint observation from distant observatories, such as Spitzer and ground telescopes \citep{Udalski2015,Street2016}. Two extra constraints on the lens mass and distance can be obtained with the use of high-resolution imaging to measure the lens flux and lens-source separation several years after the event peak \citep{Beaulieu2016}.

In every case, it is requisite to obtain strong constraints on the source star. In particular, the angular source radius is used to derive the Einstein ring radius since $\theta_E=\theta_*/\rho$. For events towards the Galactic Bulge, $\theta_*$ is generally derived from the analysis of the color-magnitude diagram (CMD) of the field, which provides an estimate of the reddening, and color-radius relations \citep{Kervella2008,Boyajian2014,Adams2018}. This method, however, cannot be used for sources in the Galactic disk because of the lower stellar density and the higher dispersion of star distances. For such events, spectroscopy is mandatory for constraining the properties and distance of the source. We note that it is often challenging to use the Gaia DR2 distance estimates \citep{Luri2018,BailerJones2018} for microlensing analysis, because of high blending in Galactic disk fields and the nuisance in the astrometric solution due to the astrometric microlensing signal \citep{Rybicki2018}. This can be quantify using the  renormalized unit weight error (RUWE) values\footnote{\url{https://gea.esac.esa.int/archive/documentation/GDR2/Gaia_archive/chap_datamodel/sec_dm_main_tables/ssec_dm_ruwe.html}}.It is also common that the microlensing source is missing from the Gaia catalogue due to the high density of stars in microlensing fields. 

The use of spectroscopy in microlensing is challenging due to the faintness of the targets and the high stellar density. However, it has been performed on several occasions. The photometric and spectroscopic observations agree at the $\sim 1 \sigma$ level for the source properties of the planetary event MOA-2010-BLG-477Lb \citep{Bachelet2012}. \citet{Bensby2013} analyzed the metallicity distribution for a sample of 58 dwarfs and sub-giants sources located in the Bulge using high-resolution spectroscopy, demonstrating a good agreement with the CMD method. With the current facilities, it is also possible to verify microlensing predictions with radial velocity follow-up of close ($\le$ 1 kpc) binary lenses. The first test done by \citet{Boisse2015} for OGLE-2017-BLG-0417 L was unsuccessful and no modulation was detected. Their null result was supported by high-resolution imaging and near-infrared spectroscopy \citep{Santerne2016}. Ultimately, \citet{Bachelet2018} resolved this puzzle by reanalyzing the microlensing lightcurves, finding a more distant lens than originally estimated. A second test of radial velocity follow-up confirmed the original binary model from a microlensing observation of the event OGLE-2009-BLG-020 L \citep{Skowron2011,Yee2016}. Spectroscopic observations obtained during the magnification of Kojima-1Lb \citep{Fukui2019} allowed for an accurate estimation of the lens system,  composed of a Super-Earth orbiting a K/M dwarf at $\sim$500 pc in the direction of the Taurus constellation. This result was confirmed by the first resolution of microlensing images ever made, using the GRAVITY instrument installed on the Very Large Telescope Interferometer (VLTI) \citep{Dong2019}. Similarly, \citet{Wyrzykowski2020} used spectroscopic observations to constrain the physical properties and distance of the giant source star in the event Gaia16aye.

The alert for the microlensing event Gaia19bld was relayed on April 18, 2019 by the Gaia Science Alerts \footnote{http://gsaweb.ast.cam.ac.uk/alerts/alert/Gaia19bld} \citep{Wyrzykowski2012,Hodgkin2013} and later recognized as a potential high-magnification event of a bright target located in the Galactic disk, in the direction of the Sagittarius Arm  ($I\sim13.5$ mag, $l=301.52358^\circ$, $b=-3.27762^\circ$) \citep{ATel12948}. Microlensing events located towards the Galactic disk are rarer than those taking place towards the Galactic Bulge \citep{Han2008,Sajadian2019,Mroz2020}, but the former offer several advantages that aid in their characterization. Indeed, they generally have larger $\theta_E$ and $t_E$, which make the previously discussed measurements somewhat simpler. 

In addition to the unusual location in the sky, Gaia19bld presented several characteristics that make  unique microlensing measurements possible, explored in three follow-up studies. The dense photometric coverage obtained from various observatories allowed for the measurement of $\theta_E$ and $\pi_E$, as described in \citet{Rybicki2021} (hereafter R21). Because the event peak magnitude was so bright, {\bf with} $H\lesssim10$ mag, it has been posssible, for the first time, to observe the microlensing images moving around the Einstein ring via interferometric measurements. The analysis of the VLT/PIONIER data leading to a direct measure of $\theta_E$ is detailed in \citet{Cassan2021}. The extreme brightening of the source has also made intensive spectroscopic follow-up studies possible, and this ultimately allowed for the precise estimation of the source properties described in this paper. The description of the data sets is presented in Section~\ref{sec:data}. Our analysis and results are detailed in Section~\ref{sec:source}. We report our conclusions in Section~\ref{sec:conclusion}.

%--------------------------------------------------------------------
\section{Observations and data reduction}
\label{sec:data}
\subsection{LCO low-resolution spectra: FLOYDS}

As the event displayed an increased brightness, spectroscopic follow-up observations were immediately scheduled. Low-resolution spectra (R$\sim$500) were obtained using the FLOYDS spectrograph, which is mounted on the Las Cumbres Observatory (LCO) 2-m telescope at Siding Spring Observatory \citep{Brown2013}. The spectral range of FLOYDS is approximately 3200 $\AA$ to 10,000 $\AA$, and a slit width of 1.2$''$ was used here. Because FLOYDS spectra suffer from fringing in the reddest parts, the spectral range was limited to $\lambda \le 7500 ~\AA$ in this work.
Our goal was to obtain {\bf a} time series of spectra to extract the spectrum of the source and the blend. We obtained three spectra around the peak of the event: on July 15, 2019 at a magnification of 55.3; on July 19, 2019 at a magnification of 36.6; and on August 1, 2019 at a magnification of 6.94. We acquired a fourth and final spectrum on February 28, 2020 at a magnification of 1.04 (proposal ID: LCO2019B-014). All spectra were reduced using the LCO FLOYDS pipeline \footnote{\url{https://github.com/LCOGT/floyds\_pipeline}} and they are presented in Figure~\ref{fig:Spectra}.  

\begin{figure}[!ht]
    \centering
    \includegraphics[width=0.4\textwidth]{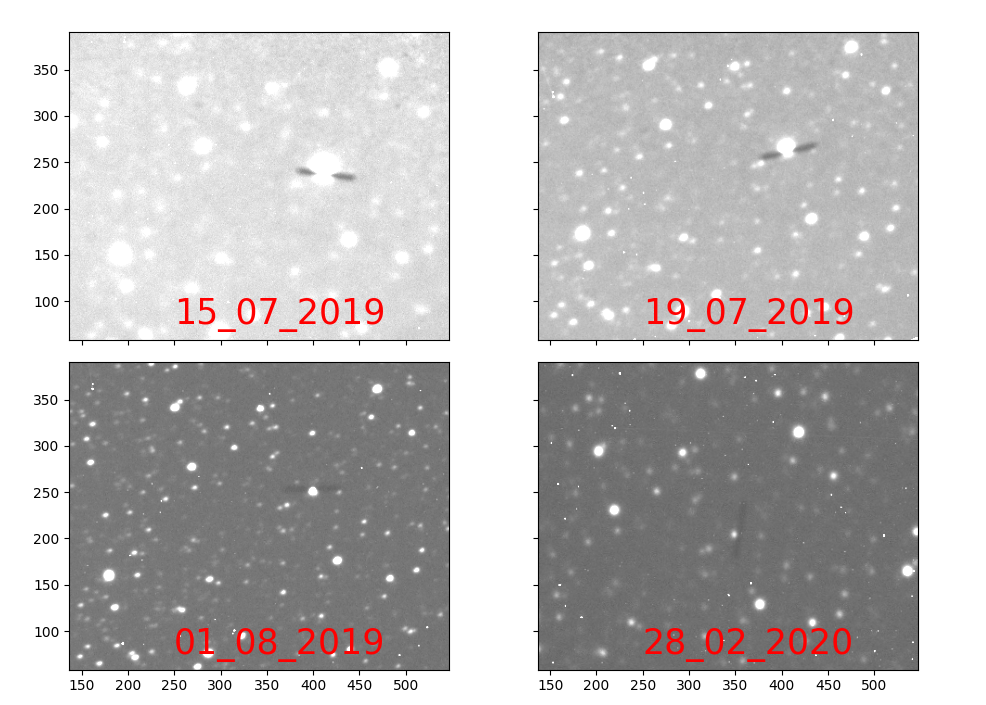}
    \caption{{Examples of guiding frames centered around Gaia19bld for each LCO FLOYDS spectra. The magnification of the source is visible. {\it Top:} Images displayed with identical scaling, so that the brighter sky background is evident in the top two images. {\it Bottom:} Decreasing target brightness resulting in sub-optimal robotic guiding.}}
    \label{fig:FLOYDS}%
\end{figure}

While the primary purpose of the FLOYDS low-resolution spectrum is to discard possible contaminants, it can confirm the spectral type of the source if the extinction is moderate towards the event \citep{Fukui2019}. Unfortunately, in the present analysis, the spectra are affected by two major flaws. Around peak magnification, when the first two spectra were obtained, the Moon was bright (at phases of 98\% and 93\%, respectively). This drastically affected the acquisition and reduction of the observations. As the event passed the peak, the last two spectra were acquired while the event was dimmer, leading to sub-optimal guiding. The effects of these two phenomena can be seen in Figure~\ref{fig:FLOYDS}. Since the field suffers from severe extinction, the measured signal in the given spectral range has a moderate signal-to-noise ratio $\rm{(S/N} \le 15)$. Therefore, we rejected these spectra from the template matching modeling described in Section~\ref{sec:templatematching}, but we note that they are in good agreement with the derived source spectrum, as can be seen in Figure~\ref{fig:Spectra}.  

\subsection{LCO high-resolution spectra: NRES}

Gaia19bld became  bright enough ($V\le12$mag) that it was possible to trigger high-resolution spectroscopy with the LCO telescopes; LCO recently deployed the Network of Robotic Echelle Spectrographs (NRES) \citep{Siverd2018}, composed of four nearly identical optical high-resolution (R$\sim$53,000) spectrographs, evenly distributed in both hemispheres. The spectrographs are fed by two sky fibres with 2.9$''$ width on sky and one simultaneous ThAr calibration fibre. The two sky fibres are attached to two 1-m telescopes at a given site, but only one telescope is selected by the scheduler for a given observation.

The NRES observations were done on two occasions: on July 15, 2019 at a magnification of 55.34 and on July 19, 2019 at a magnification of 36.13 (proposal ID: LCO2019B-014). In order to extract the wavelength calibrated spectra, we adapted the CERES pipeline \citep{Brahm2017} to reduce the NRES spectra. The pipeline starts with a bias and dark calibration of the raw images. Flat-field images are used to create the traces and the flux for each order is extracted using optimal extraction methods, as described in \citet{Horne1986}. Finally, each order is wavelength-calibrated using the ThAr calibration fibre.

\subsection{VLT/XShooter spectra}

In addition, we used the XShooter instrument \citep{Vernet2011} mounted on the ESO Very Large Telescope (VLT) which is a multi-wavelength, medium-resolution spectrograph consisted of three spectroscopic arms allowing for simultaneous observations at three wavelength ranges: UVB ($300-559.5$ nm), VIS ($559.5-1024$ nm), and NIR ($1024-2480$ nm). The resolution in each range is different due to an independent cross dispersed elements with own detector's shutters and slit masks. For UVB, VIS, and NIR ranges we were able to obtain R$\sim$5400 at slit width 1.0$''$, R$\sim$11,400 at slit width 0.7$''$, and R$\sim$8100 at slit width 0.6$''$, respectively. 
The XShooter spectrograph was used two times: on July 29, 2019 at a magnification of 8.14 (i.e., close to the peak of the microlensing event) and an airmass of 1.47; and  on November 28, 2019 at a magnification of 1.21 (i.e., close to the brightness baseline, ESO DDT proposal ID: 2103.D-5046) and an airmass of 2.15. The exposure times were 173, 207, and 2x125 seconds in the UVB, VIS, and NIR arms, respectively. We reduced the spectra with the dedicated EsoReflex\footnote{\url{https://www.eso.org/sci/software/esoreflex/}} pipeline (v. 2.9.1). For the calibration of UVB, VIS wavelengths, ThAr lamp was used, while for NIR -- a set of Ar, Hg, Ne and Xe lamps. The calibrated XShooter spectra are presented in Figure~\ref{fig:Spectra}.

\section{Properties of the source and the lens}
\label{sec:source}

\subsection{Absorption line analysis}
\label{sec:lines}

At first, the spectroscopic analysis of absorption lines for Gaia19bld is performed on two high-quality (S/N=22 and 221) XShooter VIS as well as two NRES spectra. The resolution obtained for all of these data allows for stellar parameters determination thanks to the {\it iSpec}\footnote{\url{https://www.blancocuaresma.com/s/iSpec}} framework for spectral analysis, which integrates several well-known radiative transfer codes \citep{BlancoCuaresma2014,BlancoCuaresma2019}. In our case, to determine atmospheric parameters (i.e., effective temperature $T_{\rm eff}$, surface gravity $\log g$, metallicity [M/H], microturbulence velocity $v_{\rm t}$), the SPECTRUM\footnote{\url{http://www.appstate.edu/~grayro/spectrum/spectrum.html}} code is used.

\begin{figure}[!ht]
    \centering
    \includegraphics[width=0.5\textwidth]{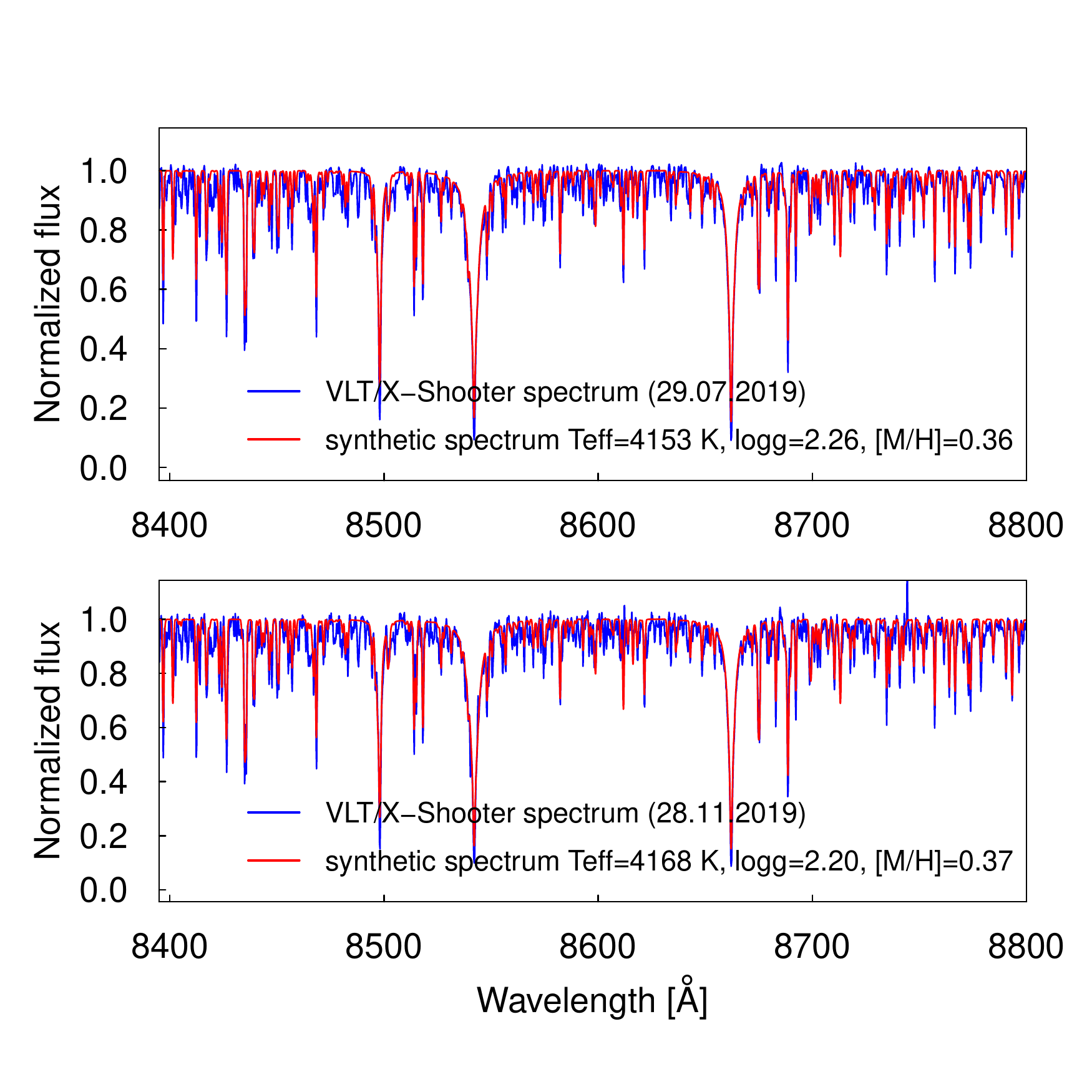}
    \caption{{Normalized XShooter spectra ({\it blue}) obtained in two epochs and synthetic spectra ({\it red}) best fits for specific parameters. The Ca~II triplet region is visible.}}
    \label{fig:linesXS}
\end{figure}

\begin{figure}[!ht]
    \centering
    \includegraphics[width=0.5\textwidth]{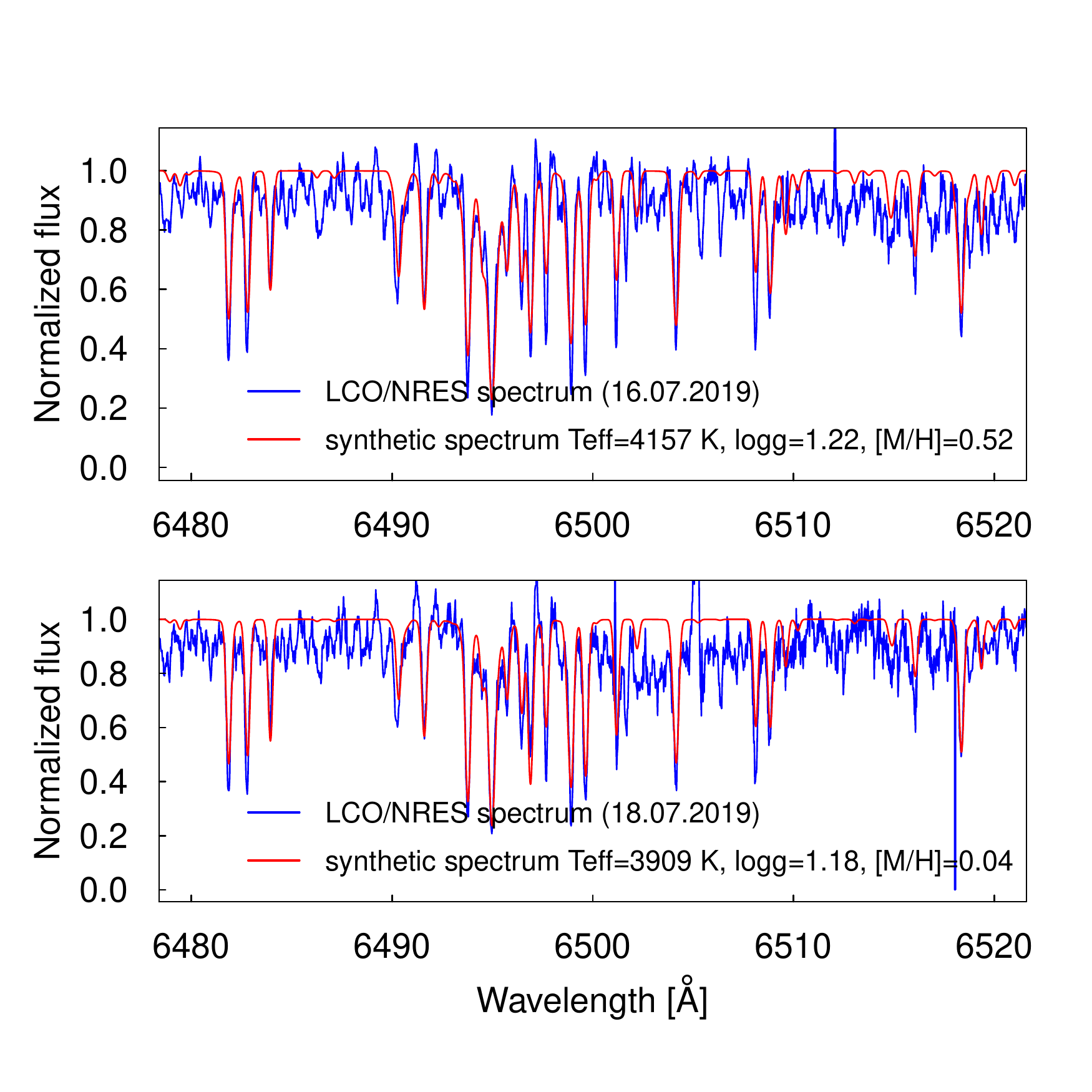}
    \caption{{Normalized NRES spectra ({\it blue}) obtained in two epochs and synthetic spectra ({\it red}) best fits for specific parameters. The Fe lines region ($6480-6520~\AA$) is visible.}}
    \label{fig:linesNRES}
\end{figure}

In order to synthesize theoretical spectra and fit them to observational data, we select prominent atomic lines (H$\alpha$, H$\beta$, Ca, Mg, Fe), a well-known grid of MARCS atmospheric models \citep{Gustafsson2008}, and solar abundances taken from \citet{Grevesse2007}. Due to the fact that the vast majority of atomic lines identified in XShooter spectra for which we do have precise laboratory data (exact wavelengths, excitation potentials, oscillator strengths values, etc.) are visible in the VIS part, we decided to focus only on this region. The best-matching synthetic spectra are fitted for parameters presented in Table~\ref{tab:modeling}. Figure~\ref{fig:linesXS} shows XShooter spectra with fitted synthetic ones around the Ca\~II triplet region, while Fig.~\ref{fig:linesNRES} shows the NRES and synthetic spectra around $6480-6520~\AA$ Fe lines. 
For all spectra, the results are the same within the uncertainties. No absorption lines from a potential second component are visible in the XShooter {\bf nor} the NRES data.

Moreover, \citet{Jennings2016} recommend the use of the $\rm{H\alpha}$ equivalent width $\rm{W(H\alpha)}$ as a potential class diagnostic for red giants because this parameter is correlated with the overpopulation of the metastable 2s level of hydrogen in non-LTE conditions. We computed $\rm{W(H\alpha)}$ for the NRES and XShooter spectra of Gaia19bld and compared them with values found in \citet{Jennings2016}. The weighted average is $\rm{W(H\alpha)}=1.13\pm0.08~\AA$ and given ${T_{\rm eff}}\sim4100$ K, the source star radius is about $40~R_{\odot}$.

\subsection{Template matching}
\label{sec:templatematching}

A second way to analyze the spectra at hand is to model the data with spectroscopic templates across the full wavelength range. This approach is complementary to the analysis presented in the previous sections. It allows us to derive an estimation of the angular radius of the source and the extinction along the line of sight. The first step is to re-arrange the data to a common wavelength basis, defined by the one used in the template model. This also considerably increases modeling speed, but introduces correlated noise in the data \citep{Carnall2017}. Then, it is also necessary to calibrate the spectra to match the predicted brightness of the event in different photometric bands. We used the magnification at the time of observation, $A(t)$, and the $I$ and Gaia-$G$ bands to estimate the parameter, $R_i$,  for each spectrum in our model. This is mandatory to derive an accurate measurement of the angular source size since $F_{\lambda}\propto \theta_*^2$. We therefore included a flux rescaling factor in the spectrum model, used a Gaussian prior, and added a contribution to the likelihood for each spectrum:
\begin{equation}
\chi^2_b=\sum_{b} {{(y_{b}-m_{{b}})^2}\over{\sigma^{^2}_{{b}}}}
,\end{equation}
where $y_b$ and $\sigma_b$ are the observed fluxes and errors in the band $b$ at the time of the spectrum acquisition. We assumed a conservative $\sigma=0.1$ mag for all photometric measurements.

\begin{figure*}[!ht]
    \centering
    \includegraphics[width=0.9\textwidth]{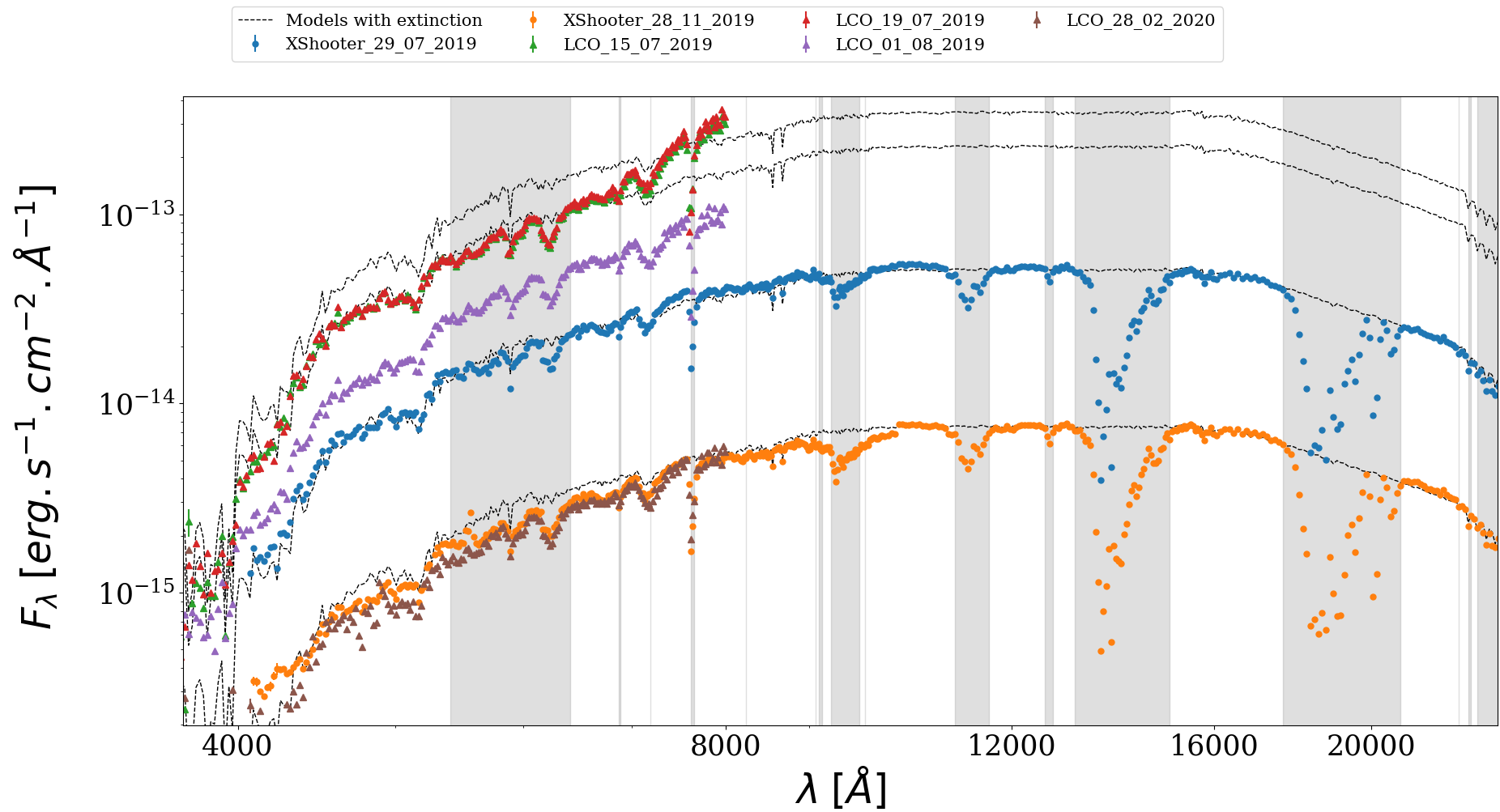}
    \caption{Four LCO FLOYDS and two XShooter spectra clearly indicating the magnification of the source. Dash lines indicate the best fit model (model A, displayed for four different magnification levels) from the template-matching analysis, while the grey regions correspond to telluric lines, with absorption $\ge2$\%. The LCO FLOYDS spectra confirm the spectral type but were not used for the modeling. The observed spectra have been scaled to the observed magnitude at the time of acquisition (i.e., they represent the spectra as seen at the top of the atmosphere).}
    \label{fig:Spectra}%
\end{figure*}

To model the spectra, we used the \textit{pysynphot} software \citep{STScI2013} for generating the spectral templates from \citet{Kurucz1993}. This allowed us to estimate the effective temperature, $\rm{T_{eff}}$, the surface gravity, $\rm{logg,}$ and the metallicity, $\rm{[Fe/H],}$ of the star. To match the model to the data, the angular source radius $\theta_*$ is also required and used to scale the flux. Finally, we used the relations from \citet{Cardelli1989} to estimate the extinction towards the source, parameterized by $A_V$ and $R_V$. The microlensing model is defined as:
\begin{equation}
f_\lambda(t)=f_{s,\lambda}A(t)+f_{b,\lambda}
\label{eq:model}
,\end{equation}
where $f_{s,\lambda}$ and $f_{b,\lambda}$ are the source and blend fluxes at wavelength $\lambda$. We used the magnification from the photometric models detailed in R21. We consider the errors of the spectra to potentially be underestimated and add to the model the parameter $k$ for each spectrum:
\begin{equation}
\sigma^{'}_{\lambda}=\sqrt{k^2\sigma_{\lambda}^2}
,\end{equation}
with $\sigma^{'}_{\lambda}$ and $\sigma_{\lambda}$ the new and original errors at the wavelength $\lambda$. The terms k take account of the flux calibration errors and can be expected to be on the order of $\sim{{\sigma}\over{\sqrt{N}}} \rm{S/N}$, where the S/N is the mean signal-to-noise ratio of the spectra. The two XShooter spectra have S/N$\sim$1700 and and S/N$\sim$650 (after binning) and thus we should expect $k_1\sim$120 and $k_2\sim$40, respectively. Ultimately, we model the data by maximizing the likelihood $L$:
\begin{equation}
L=-0.5\sum_{\lambda}\bigg( {{(y_{\lambda}-m_{{\lambda}})^2}\over{\sigma^{'2}_{{\lambda}}}}+\chi^2_b+\ln(2\pi\sigma^{'}_{\lambda})\bigg)
.\end{equation}
To explore the posterior distribution, we used the \textit{emcee} package \citep{Foreman2013}. We rejected points with atmospheric extinction higher than 2\% and used the telluric lines defined in \citet{Moehler2014}. We identify a single dominant minima, noted A in Table~\ref{tab:modeling}  and visible in the Figure~\ref{fig:Spectra}. It can be seen in Figure~\ref{fig:Spectra} that the second XShooter spectrum present a slight discontinuity starting around 10,000 $\AA$. A close look at the data reveals a small offset  between the calibration of the VIS and NIR arms, which we found to be $\epsilon\sim0.87$ (a multiplicative offset). The cause of this offset can be due to the observation at low airmass (2.15), where the hardware and software corrections can underperform\footnote{\url{https://www.eso.org/sci/facilities/paranal/instruments/xshooter/doc/VLT-MAN-ESO-14650-4942_P104v1.pdf}}. After the correction of this offset, we ran a second round of modeling that converges to the solution B. Finally, we found an extra solution, noted C, by entirely removing  the  second X-Shooter spectrum. All solutions indicate that the source is a red giant, with slightly different spectral types and extinction properties for each individual solution.  The absorption, $A_V$, derived in model A, B, and C are in agreement with the independent estimations, that is, $2.01\pm0.03$ and $2.34\pm0.04$ from \citet{Schlafly2011} and \citet{Schlegel1998} (both assuming $R_V=3.1$\footnote{https://irsa.ipac.caltech.edu/applications/DUST/}), and $1.88\pm0.35$ from \citet{Anders2019}. To compare these results with the photometric measurements, we computed the magnitude of the source in several bands for all models. As summarized in Table~\ref{tab:abpredicted}, the flux at the event location is almost solely due to the source. Using the zero-point correction from \citet{Blanton2007} and \citet{Weiler2018}, we found the source magnitude to be $G=14.8\pm0.2$ mag, $V=15.7\pm0.2$ mag, and $I=13.4\pm0.2$ mag. This is in excellent agreement with the measurements of R21 with $G=14.8$ mag, $V=15.9$ mag, and $I=13.5$ mag, and it is in good agreement with archival data\footnote{\url{https://vizier.u-strasbg.fr/viz-bin/VizieR-4}}. All models predict very similar magnitudes in all bands and are in agreement at the $1\sigma$ level. We note, however, that models B and C are in better agreement with the archival near-infrared data from \citet{Cutri2003}. 

\begin{table*}
  \renewcommand{\arraystretch}{1.5}
\centering
\caption{Summary of the derived parameters for the source of Gaia19bld event. Averaged solutions of line fitting are presented. The three minima A, B, and C from the template matching method discussed in the text are presented. The physical parameters $\theta_E$, $M_l$, $D_s$, and $D_l$ have been estimated using the $\rho$ and $\pi_E$ {\bf parameters} from the light curve modeling presented in R21.} \label{tab:modeling}
\begin{tabular}{lcccc}
\hline
\hline
Parameter          & Line fitting  & \multicolumn{3}{c}{Template Matching} \\
&&A&B&C\\
\hline
$T_{\rm eff}$ [K]  & $4159\pm139$  &$4097^{+32}_{-29}$& $4052^{+27}_{-28}$ & $4091^{+31}_{-31}$\\
${\log g}$    & $1.89\pm0.42$ &$1.48^{+0.15}_{-0.16}$& $1.27^{+0.15}_{-0.17}$ & $1.49^{+0.15}_{-0.16}$ \\
${\rm [M/H]}$  [dex]    & $0.42\pm0.20$ &$0.295^{+0.053}_{-0.062}$ & $0.219^{+0.050}_{-0.054}$ & $0.289^{+0.055}_{-0.062}$  \\
$v_{\rm t}$ [km/s] & $2.04\pm0.58$ & -- & -- & --  \\
$A_{\rm v}$ [mag]  &    --         &$2.322^{+0.075}_{-0.072}$ & $2.153^{+0.064}_{-0.070}$ & $2.214^{+0.074}_{-0.076}$ \\
$R_{\rm v}$   &    --         &$3.40^{+0.11}_{-0.10}$&$3.13^{+0.072}_{-0.069}$ & $3.117^{+0.089}_{-0.085}$  \\
${\rm \theta_*}~[\mu as]$ &   --   &$24.16^{+0.39}_{-0.40}$&$23.20^{+0.38}_{-0.38}$& $23.11^{+0.53}_{-0.55}$ \\
$k_1$ & -- & $117.0^{+7.0}_{-6.3}$ & $103.0^{+4.4}_{-4.0}$ &$101.5^{+4.3}_{-3.8}$ \\
$k_2$ & -- & $68.6^{+4.0}_{-3.8}$ & $48.1^{+2.0}_{-1.9}$ & --\\

L&--&21430&21582&10595\\
\hline
$\theta_E ~\rm{[\mu as]}$ & -- & $754^{+13}_{-13}$ & $724 ^{+12}_{-12}$ & $721^{+17}_{-18}$\\
$M_l ~\rm{[M_\odot]}$ & -- & $1.126^{+0.027}_{-0.026}$ &$1.081^{+0.026}_{-0.025}$ & $1.076^{+0.031}_{-0.031}$\\
$D_s$ [kpc] & -- & $8.4^{+1.3}_{-1.8}$ & $9.3^{+1.5}_{-1.9}$ &$8.7^{+1.4}_{-1.9}$\\ 
$D_l$ [kpc] & -- & $5.50^{+0.56}_{-0.82}$ & $5.97^{+0.59}_{-0.84}$ & $5.74^{+0.58}_{-0.86}$\\ 
\hline
\hline
\end{tabular}
\end{table*}

%\begin{figure*}[!ht]
%    \centering
%    \includegraphics[width=0.9\textwidth]{MCMC_Gaia19bld_new.png}
%    \caption{{Posterior distribution of the template matching parameters.}}
%    \label{fig:MCMC}%
%\end{figure*}

\subsection{ Einstein ring radius, source distance, and lens properties}
The photometric analysis presented in R21 was converged to $\rho=0.03198\pm0.00016$ and $\pi_E=0.0815\pm{0.0014;}$ we used these values to compute the values of $\theta_E$ and $M_l$ presented in Table~\ref{tab:modeling}. These values are in agreement with the independent measurement $\theta_E = 0.765\pm0.004$ mas from interferometry (C21), especially for model A.

\begin{figure}[!ht]
%\label{fig:sourcedistance}
    \centering
    \includegraphics[width=0.4\textwidth]{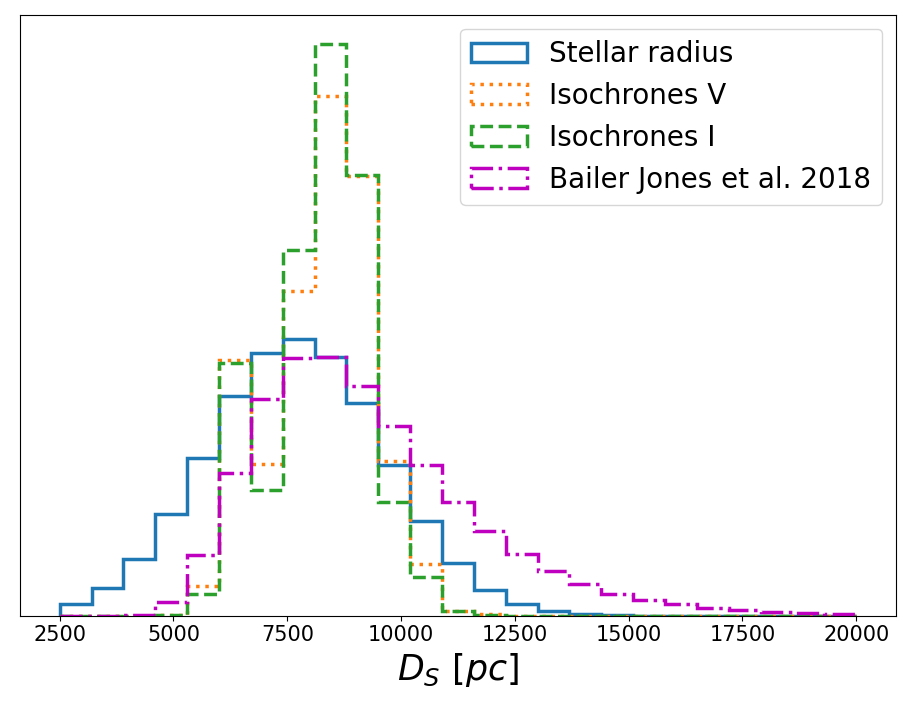}
    \caption{Distributions of the distance of the source from the various methods detailed in the text (model A).}
    \label{fig:distance}%
\end{figure}

One of the difficulties of microlensing studies is the estimation of the source distance, $D_s$. This is especially true for fields outside of the Galactic Bulge, where the extinction is not well known. However, the angular radius of the source and fundamental stellar properties derived in previous sections allow us to estimate the source distance with four different methods. Here, we detail the source distance estimate made for model A, however, all the results can be found in Table~\ref{tab:modeling}. First, it is clear from the fit parameters that the source in Gaia19bld is a red giant. Therefore, using the empirical relations presented in Section~\ref{sec:lines}, and from \citet{Berger2018} and \citet{Alonso2000}, we can assume the physical radius of the source to be $R_*=40 \pm 10~  R_\odot$. We assumed a relatively low precision on the stellar radius due to the intrinsic scatter in the $T_{\rm eff}$-radius relation for the red giant population, as well as the interpolation methods used to estimate these radii. We note, however, that using the stellar parameters of the source with the \textit{PARSEC} isochrones \citep{Bressan2012} returns a similar value {\bf of} $R_*=44 \pm 3~  R_\odot$. This estimation, associated with the posterior distribution of the angular source radius, led to a direct measurement of the source distance: $Ds=7.7_{-1.9}^{+1.9}$ kpc. Secondly, using the \textit{PARSEC} isochrones \citep{Bressan2012} \footnote{http://stev.oapd.inaf.it/cgi-bin/cmd}, with a fixed age of 1 Gyr, and the derived source parameters, it is possible to estimate the absolute magnitude, $M_V\sim -1.0$ mag  and $M_I\sim-2.5$ mag, of the source. Coupled with the apparent magnitudes of the source, $V=15.85$ mag and $I=13.52$ mag, and the absorption law derived from spectral modeling, we obtain two extra source distance estimates: $Ds=8.4_{-1.6}^{+0.9}$ kpc and $Ds=8.4_{-1.5}^{+0.8}$ kpc. Finally, we use the expressions from \citet{BailerJones2018} for a fourth independent distance estimate of the source $Ds=8.9_{-1.8}^{+2.7}$ kpc. The four resulting distributions for the model A can be seen in Figure~\ref{fig:distance}. Combining all the distributions gives a distance estimate of $Ds=8.4^{+1.4}_{-1.8}$ kpc. 

According to Equation~\ref{eq:model}, it is possible to extract the source and blend the spectra if at least two measurements were obtained at two different magnifications. We did not find any significant blend light in this case (after the correction of the offset of the second XShooter spectrum). It is also possible to estimate the covariance matrix as:
\begin{equation}
\mathbf{C} = \mathbf{H^{-1}}
,\end{equation}
where $\mathbf{H}$ is the Hessian matrix which is equal to:
\begin{equation}
\mathbf{H} = \mathbf{M^{T}M}
,\end{equation}
and
\begin{equation}
\mathbf{M} = \biggl ( \begin{matrix} \mathbf{A}&\mathbf{B}\\ \mathbf{B}&\mathbf{C}\end{matrix} \biggr )
.\end{equation}
Here, {\bf A}, {\bf B}, and {\bf C} are diagonal square matrices of dimension $N_\lambda \times N_\lambda$ (where $N_\lambda$ is the total number of observed wavelengths), assuming the $N_s$ spectra have been observed at different magnifications, $A_s$, over the course of the microlensing event. The diagonal of matrices {\bf A}, {\bf B}, and {\bf C} can be written as:
\begin{equation}
A_{\lambda} = \sum_{i=1}^{N_s} {{A_i^2}\over{\sigma_{\lambda,i}^2}} ~ ; ~ B_{\lambda} = \sum_{i=1}^{N_s} {{A_i}\over{\sigma_{\lambda,i}^2}} ~;~
C_{\lambda} = \sum_{i=1}^{N_s} {{1}\over{\sigma_{\lambda,i}^2}}
.\end{equation}

Using the two XShooter spectra, we estimate the error on the measured source and blend fluxes to be $\sigma_{f_{s,\lambda}}\sim3. 10^{-16} ~\rm{erg.s^{-1}.cm^{-2}.\AA^{-1}}$ and $\sigma_{f_{b,\lambda}}\sim6. 10^{-16}~\rm{erg.s^{-1}.cm^{-2}.\AA^{-1}}$, respectively. The latter value places a conservative upper limit on the detectable flux emitted by the blend and, ultimately, the lens. This flux density upper limit is equivalent to the magnitude's uppper limit of $\sim V\le17$ mag. As presented in Table~\ref{tab:abpredicted}, the main sequence lens scenario predicts $V\sim20$ mag, which is much fainter than the previous limit. Therefore, the current spectroscopic data does not bring additional constraints on the nature of the lens. With a predicted blend flux ratio in the V-band, namely, $g_v\sim2\%$, we note that this analysis is consistent with the absence of blend light reported in the photometric results presented in R21.

\begin{table*}[h!]
  \renewcommand{\arraystretch}{1.5}
\centering

\caption{ Predicted AB magnitudes for the source, the main-sequence lens scenario as well as archival measurements at the event location. The number in brackets indicates the standard uncertainties. The archival magnitudes have been corrected to the AB system, using \citet{Blanton2007} and \citet{Weiler2018}.} \label{tab:abpredicted}
\begin{tabular}{l|ccccc}

\hline
\hline
Filter & \multicolumn{3}{c}{Source} &  Main Sequence Lens & Archival data  \\ 
&A&B&C \\
 \hline

G &$14.88(0.08)$&$14.89(0.07)$&$14.86(0.06)$&$19.9(0.5)$& $14.897(0.003)$ \citep{Gaia2016,Gaia2020} \\
U &$21.2(0.2)$&$21.3(0.2)$&$21.3(0.2)$& $23.0(0.5)$& *\\
B &$17.7(0.1)$& $17.7(0.1)$& $17.7(0.1)$& $21.5(0.6)$&17.52 \citep{Girard2011} \\
V &$15.8(0.1)$&$15.8(0.1)$&$15.8(0.1)$& $20.3(0.5)$&  15.99 \citep{Girard2011} \\ 
R &$14.78(0.09)$&$14.79(0.08)$&$14.8(0.1)$&$19.7(0.5)$& 14.6 \citep{Zacharias2005}  \\
I &$13.92(0.07)$&$13.94(0.07)$&$13.93(0.08)$& $19.2(0.5)$& $13.93(0.02)$ \citep{DENIS2005} \\
u &$21.1(0.2)$&$21.2(0.2)$&$21.2(0.2)$& $23.1(0.6)$& * \\
g &$17.0(0.1)$&$17.1(0.1)$&$17.1(0.1)$& $21.1(0.6)$&* \\
r &$15.1(0.1)$&$15.1(0.1)$&$15.1(0.1)$& $19.9(0.5)$&* \\
i &$14.22(0.08)$&$14.24(0.07)$&$14.23(0.09)$& $19.4(0.5)$&* \\
z &$13.50(0.07)$&$13.54(0.06)$&$13.52(0.08)$& $19.0(0.5)$&* \\
J&$12.63(0.05)$&$12.69(0.05)$&$12.67(0.06)$&$18.6(0.5)$&$12.76(0.03)$ \citep{Cutri2003}\\
H&$12.12(0.04)$&$12.18(0.04)$&$12.18(0.04)$&$18.5(0.5)$&$12.25(0.02)$ \citep{Cutri2003} \\
K &$12.29(0.04)$&$12.36(0.04)$&$12.37(0.04)$& $18.8(0.5)$&$12.36(0.02)$\citep{Cutri2003}\\

\hline
\hline
\end{tabular}
\end{table*}

%We illustrate this situation bin the Figure ???. We assumed three temperature for white dwarfWe assumed the lens suffer the same ammount of extinction as the source.

%\begin{figure}[!ht]
%\label{fig:sourcedistance}

%    \centering
%    \includegraphics[width=0.4\textwidth]{Gaia19bld_lenses.png}
%    \caption{{Distance of the source from the various methods detailled in the text.}}
%    \label{fig:distance}%
%\end{figure}
\section{Conclusions}
\label{sec:conclusion}
In this work, we present spectroscopic follow-up studies of the microlensing event Gaia19bld. We collected several spectra on different instruments over the course of the event. For the first time, we performed a joint analysis of the absorption lines of the XShooter and NRES high resolution, as well as a template-matching modeling. The spectra lines and template matching analysis converge to similar solutions. The source is a red giant ($T_{\rm eff}\sim4100$ K, $\log~g\sim1.5$, ${\rm [M/H]}\sim0.3$ and $\theta_*\sim~24 \mu as$), located at $\sim$ 8.4 kpc from the Earth. We did not measure any significant blend light and, therefore, there is no detection of the lens in the spectroscopic data. Indeed, the combination of the measurement of $\theta_*$ and the parameters extracted from the light curve (R21) leads the a lens mass $\sim1.1 M_\odot$ located at $D_l\sim5.5$ kpc. At this distance, a main sequence lens would be too faint to have been detected in spectroscopic (and photometric) data because it is much fainter ($V\sim20$ mag) than our detection limit ($V\le$ 17 mag). 

This work demonstrates the potential of the spectroscopic follow-up of the microlensing event. It allows for the precise characterization of the source star stellar parameters and its angular radius, as well as the extinction along the line of sight. This is especially useful for events in the Galactic disk, where the distance to the source and the extinction are not well known. It is expected that the methods described in this work will be used routinely in the era of the new generation of all-sky surveys currently under development. In particular, the Legacy Survey of Space and Time \citep{LSST2009} will detect thousands of events in the Galactic disk every year  \citep{Sajadian2019}. This will require similar spectroscopic monitoring in order to better characterize their properties and ultimately improve our understanding of faint objects throughout the entire Milky Way.

\textit{Software: Astropy \citep{Astropy2018}, emcee \citep{Foreman2013}, Spyctres\footnote{https://github.com/ebachelet/Spyctres}, pyLIMA \citep{Bachelet2017}, pysynphot \citep{STScI2013}}

\begin{acknowledgements}
EB and RS gratefully acknowledge support from NASA grant 80NSSC19K0291.
YT acknowledges the support of DFG priority program SPP 1992 "Exploring the Diversity of Extrasolar Planets" (TS 356/3-1).
This work is supported by Polish NCN grants: Daina No. 2017/27/L/ST9/03221, Preludium No. 2017/25/N/ST9/01253, Harmonia No. 2018/30/M/ST9/00311 and MNiSW grant DIR/WK/2018/12 as well as European Commission's Horizon2020 OPTICON grant No. 730890.

This paper uses data collected with ESO/VLT/XSHOOTER instrument allocated via DDT programme No. 2103.D-5046. We thank the ESO staff for their support. 

This work has made use of data from the European Space Agency (ESA) mission
{\it Gaia} (\url{https://www.cosmos.esa.int/gaia}), processed by the {\it Gaia}
Data Processing and Analysis Consortium (DPAC,
\url{https://www.cosmos.esa.int/web/gaia/dpac/consortium}). Funding for the DPAC
has been provided by national institutions, in particular the institutions
participating in the {\it Gaia} Multilateral Agreement.

This research has made use of the VizieR catalogue access tool, CDS,
 Strasbourg, France (DOI : 10.26093/cds/vizier). The original description 
 of the VizieR service was published in 2000, A\&AS 143, 23
\end{acknowledgements}

\bibliographystyle{aa}  % needs package natbib

\bibliography{microlensing.bib}

\end{document}